# Molecule optimization via multi-objective evolutionary in implicit chemical space

Xin Xia[1], Yansen Su[1]*, Chunhou Zheng[1], Xiangxiang Zeng[2]*

1 The Key Laboratory of Intelligent Computing and Signal Processing of Ministry of Education, School of Artificial Intelligence, Anhui University, Hefei 230601, China.

2 College of Computer Science and Electronic Engineering, Hunan University

Email: suyansen@ahu.edu.cn, xzeng@foxmail.com

## Abstract

Machine learning methods have been used to accelerate the molecule optimization process. However, efficient search for optimized molecules satisfying several properties with scarce labeled data remains a challenge for machine learning molecule optimization. In this study, we propose MOMO, a multi-objective molecule optimization framework to address the challenge by combining learning of chemical knowledge with Pareto-based multi-objective evolutionary search. To learn chemistry, it employs a self-supervised codec to construct an implicit chemical space and acquire the continues representation of molecules. To explore the established chemical space, MOMO uses multi-objective evolution to comprehensively and efficiently search for similar molecules with multiple desirable properties. We demonstrate the high performance of MOMO on four multi-objective property and similarity optimization tasks, and illustrate the search capability of MOMO through case studies. Remarkably, our approach significantly outperforms previous approaches in optimizing three objectives simultaneously. The results show the optimization capability of MOMO, suggesting to improve the success rate of lead molecule optimization.

**Introduction**

Drug discovery aims to find molecules with biological activity, high drug-likeness, safety simultaneously, which is a time-consuming and expensive process with a high failure rate[1, 2]. Before a molecule can be developed into a laboratory-approved drug, it must be screened and optimized for various properties, and pass multiple phases of preclinical and clinical trials[3]. During the process, molecule optimization (MO) is a crucial part to acquiring the optimum drug[4]. The task of MO is to optimize the lead molecule to generate similar molecules with desired properties[5]. How to harmoniously optimize multiple properties and similarity of molecules is a critical challenge that needs to be addressed.

Traditional approaches to MO are mainly based on virtual screening or experimental trial-and-error[6-8], which are time-consuming and energy-intensive. As reported[9], the number of molecules in chemical space is estimated to be between $10^{30}$ and $10^{60}$, so artificially exploring the chemical space is extremely limited. Machine learning approaches and techniques have developed rapidly and widely used in the MO field to automatically generate new molecules[10-12], which explore the chemical space more broadly, and accelerate the drug discovery procedure. Among them, learning-based data-driven approaches capture chemical knowledge to meaningfully optimize molecules, increasing the validity and success rate of the optimized molecules[13, 14].

Although several excellent molecule design models can used to optimize molecules[15-18], they generally emphasize property optimization, the similarity between the optimized molecule and the original molecule needs additional improvement. For example, the model proposed by Tiago et al.[18] generates novel molecules with high properties. However, as shown in the results of the constrained PlogP optimization task, the average properties of the optimized molecules decrease significantly as the similarity constraint increases, and the optimized molecules have lower similarity values than the baselines. Therefore, designing computational models to optimize both properties and similarity is necessary. Compared to molecule design models, MO models focus more on the optimization of similarity. The rise of advances in deep learning (DL) models motivate their application to MO, including VAE[19-23], RNN[24-27], GAN[28-30] and GNN[5, 31-33], etc. DL-based approaches transform discrete molecular representations into continuous latent representations, and automatically generate new molecules by performing simple operations in the latent space. These approaches require large amounts of training data that are not easily accessible. Moreover, the objective function needs to be specifically designed, as molecular properties are non-differentiable and cannot be adapted to directly train the model. In comparison, reinforcement learning (RL) methods can directly apply complex and non-differentiable molecular properties as rewards, and generate molecules through the designed actions and rewards. Existing RL-based approaches normally model MO as a Markov decision process (MDP), and progressively modify molecules by adding characters on SMILES or adding/deleting atom or bond on molecule graph[34-37]. In fact, searching in discrete chemical space via stepwise modification of molecules is inefficient. Evolutionary algorithms (EAs) also directly use molecular properties as objectives, and generate new molecules by defining crossover and/or mutation operations based on different molecular representations[38-41]. Compared with modifying molecules by stepwise in RL, EA directly obtains the final state of molecules, which is highly nonlinear, easily modifiable and parallelizable[42]. Though flexible and straightforward, both RL and EA methods lack a priori knowledge of chemistry, leading to low efficiency when searching in discrete spaces. In recent work, several hybrid methods combine deep learning with EA[43] or RL[44] to solve the efficiency problems and improve the performance to some extent.

The aforementioned MO models provide some improvement of molecular properties and similarity. However, most existing approaches either optimize a single objective with similarity constraint[21, 22, 25] or aggregate multiple objectives into a single objective optimization problem[13, 14, 26]. Combining multiple objectives into a single objective via weighted summation or product has the following limitations[45]: 1. The weights used to trade off the relative importance of multiple objectives are hard to determine and must be adjusted experimentally. 2. An optimum solution to the scalarized objective with a preference (weights) is just a point on (or near) the Pareto-front, the solutions for various preferences require repeating the optimization process waste plenty of computational resources. 3. Scalarization approaches by weighted summation can only approximate convex Pareto-front, which are irregular and possibly intermittent for MO. 4. The non-uniformity of magnitudes among the objectives may lead to poor robustness. Therefore, it is difficult to fully explore the chemical space with a single-objective optimization approach. Recently, multi-objective evolutionary algorithms (MOEAs) have been better developed to simultaneously optimize multiple conflicting objectives[46-50]. MO is naturally a multi-objective optimization problem[51], so it is worthwhile to use MOEAs to optimize multiple properties and similarity of molecules. Despite this, difficult-to-access label data and low search efficiency remain major issues in machine learning based MO[52].

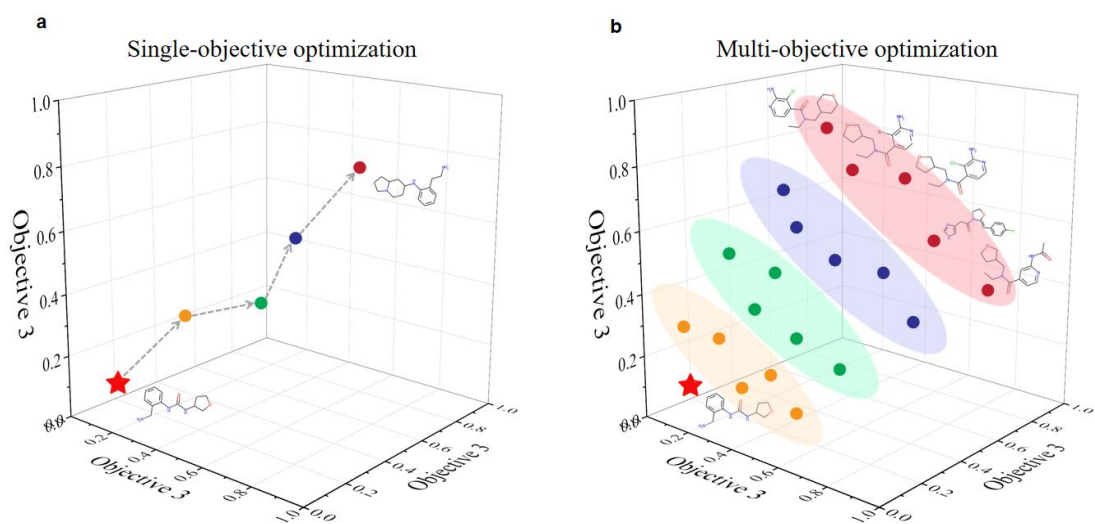

**Fig.1. Optimization procedure for single- and multi-objective molecule optimization.** The red pentagram is the lead molecule, and the dots with different colors indicate the molecules generate in the search process. **(a)** Single-objective optimization, where only one solution is acquired in an optimization procedure. **(b)** Multi-objective optimization, which can find a set of solutions that reveal different trade-offs between objectives in an optimization procedure, and explore the chemical space more comprehensively.

In this study, we propose the multi-objective molecule optimization (MOMO) framework via the combination of learning and evolution, which is a flexible and efficient method to optimizing multiple properties and similarity not relying on labeled data (**Fig. 1**). MOMO employs a deep learning model to construct an implicit chemical space to be explored by learning chemistry from a large set of existing molecules, and combines a Pareto-based multi-objective evolutionary optimization to purposefully explore a set of optimal molecules for different trade-offs between multiple objectives in the implicit chemical space. Compared with state-of-the-art methods, the advantages of the MOMO model are as follows: (1) The use of multi-objective evolutionary algorithm prompts MOMO to generate more abundant and diverse molecules that satisfy various property requirements and explore the chemical space more comprehensively. (2) Learning about chemistry improves search efficiency. Evolution in implicit chemical space can explore molecules more efficiently and smoothly, without artificially designed rules, and disentangle from the constraints of discrete molecular representations. (3) MOMO reduces the dependence on labeled data by self-supervised training of codecs and evolving molecules. Experimental results show that the MOMO method achieves state-of-the-art performance on all four tasks. In particular, in multi-property optimization tasks, MOMO generates similar molecules with several higher properties than the lead molecule.

## Results

**MOMO framework**

The task of MO requires the balanced optimization of multiple conflicting properties simultaneously. Most existing models address this problem by combining multiple objectives into a single objective, which relies on assumptions about the trade-offs between the objectives[53]. For actually, the optimal solution for a preference is only one of the solutions on the Pareto front. Finding solutions for different preferences must repeat the experiment, which consumes computational resources. In contrast, Pareto-based EAs identify a set of solutions for various trade-offs of objectives at a time, allowing the expert to select the optimal molecule considering different relative importance of the objectives. However, some challenges are recognized that need to be addressed for the use of EAs in MO. First, EAs search for solutions without prior knowledge. However, for MO, blindly evolving molecules in the absence of chemical knowledge is a waste of computational resources and leads to inefficiencies. Second, how to represent discrete molecules for evolution is also a challenge, which further determines the computational complexity and exploration space. For example, graph- and sequence-based EAs that explore discrete spaces need to satisfy various constraints of chemical rules[54].

To this end, we propose MOMO framework, which decouples the representation learning and search part of the molecule, uses a deep learning model to capture the chemistry knowledge and construct the implicit chemical space to be explored, further obtain the continuous representation of molecules, and applies a Pareto-based MOEA to search for a set of molecules to satisfy multiple properties and similarity in the implicit chemical space. The main framework of the model is shown in **Fig. 2**, and the optimization processes are as follows.

(1) An encoder-decoder module is pre-trained on a large public molecule database to learn chemistry and build an implicit chemical space.

(2) The lead molecule is embedded into the implicit space and the latent representation of the molecule is obtained through the encoder.

(3) If it is the first generation, MOMO constructs an initial population by adding Gaussian noises to perturb the latent vector of the input molecule. The individual in the population is an embedding vector of molecules (i.e., a solution for the optimized molecules). Alternatively, if it is not the first generation, the molecules in the new population from the previous generation are encoded.

(4) Evolutionary operations (selection, crossover, and variation) are conducted on latent vectors in population to generate new vectors of offspring.

(5) The latent vectors in the parental population and offspring are merged and decoded into molecule space by the pre-trained decoder.

(6) The values of multiple objectives of these evolved molecules are acquired by the property evaluator in the molecule space[55-57].

(7) According to the multi-objective values, the partial order of the molecules is defined in terms of the non-domination rank and the reference point mechanism. To avoid local optimality in the evolution, we design the acceptance probability to accept high-quality molecules into the next generation with a dynamic probability that varies with iteration. Molecules are then selected into the new population via the partial order and the acceptance probability.

(8) Steps (3)-(7) are iterated to search for molecules in the latent space and the last generation population are finally returned. The molecules on the Pareto-front in the final population are the optimized molecules.

The main advantages of the MOMO framework are explained below. In step (1), the pre-trained codec learns the intrinsic relationship between molecules to construct a continuous implicit space. The learned chemical knowledge is used to improve the validity of the molecules generated in the evolution to speed up the search process. In step (2), continuous vectors are employed as coding in the evolutionary procedure rather than discrete molecules, which makes the evolutionary operation more efficient and smooth. In step (3), we take into account the problem of scarce and expensive label data in practical, generate the initial population by perturbing the embedding vector of the lead molecule without additional molecules. In steps (4) to (6), evolutionary operations are performed in the implicit space to smoothly explore the latent space, while evaluation are performed in molecule space to ensure the evaluation accuracy of properties and similarity. The Pareto-based optimization approach used in step (7) explores the chemical space comprehensively by combining the non-domination rank and reference point mechanism. The designed dynamic acceptance probability further regulates the trade-off between convergence and diversity during the search. This means that more diverse molecules are explored at early evolutionary stages, while more high-quality molecules are selected at later stages. Ultimately, we obtain a set of optimal molecules with various trade-offs of multiple targets, increasing the success probability of MO. The detailed MOMO framework is described in the Methods section.

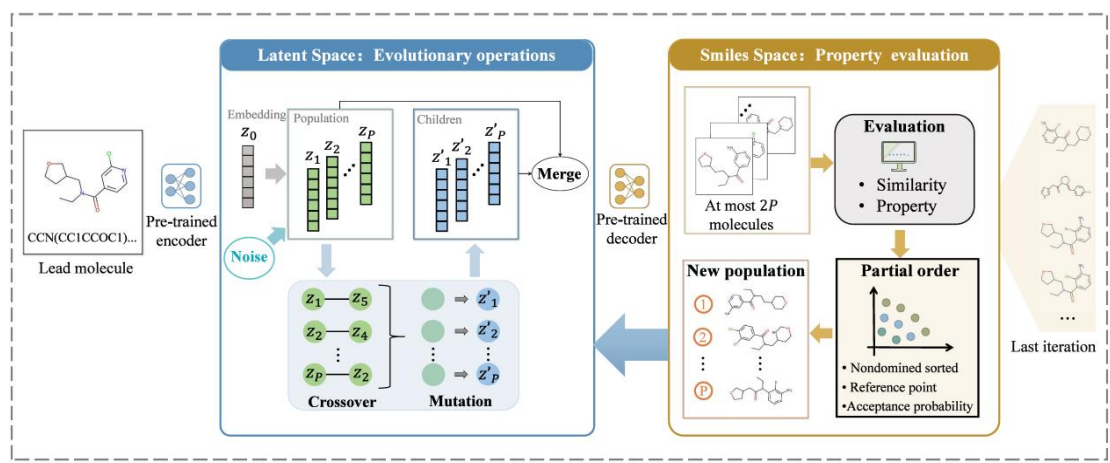

**Fig. 2 The framework of MOMO.** MOMO optimizes the target properties of lead molecules by evolving the molecular latent representation in an implicit chemical space, evaluating properties in the molecule space, and preserving the desired molecules. The implicit chemical space is constructed by a pre-trained encoder-decoder module. Given a candidate embedding $z_0$, the initial population is acquired by adding Gaussian noises to perturb the vector, then the new molecules are generated by implementing evolutionary operations (selection, crossover and mutation). After evaluating the properties and similarity of the corresponding decoded molecules, the desired molecules are composed into the next population based on their partial order according to the multiple objectives and the designed dynamic acceptance probability.

**Benchmark evaluation of MOMO**

To demonstrate the effectiveness of MOMO, we benchmark the performance on four multi-objective MO tasks, which are set up according to widely used benchmark properties in MO domains and reported papers[13, 26].

The target properties in experiments include the drug likeness (QED)[58], Penalized-logP (PlogP)[21], Dopamine Receptor D2 (DRD2)[59] and Tanimoto similarity (Similarity)[60] between the optimized molecule and the lead molecule. The four tasks are designed by combining these properties and similarity as follows:

(1) Task 1: QED and Similarity

In this task, the objective is to optimize the lead molecule to $\text{QED} \geq 0.9$ and $\text{Similarity} \geq 0.4$.

(2) Task 2: PlogP and Similarity

In this task, the aim is to maximize the improvement of PlogP while satisfying the similarity threshold. The improvement of PlogP is denoted as PlogP_imp, and the threshold in this task is set to either 0.4 or 0.6.

(3) Task 3: QED, PlogP and Similarity

This task aims to optimize molecules to $\text{QED} \geq 0.85$, $\text{PlogP\_imp} \geq 3$ and $\text{Similarity} \geq 0.3$.

(4) Task 4: QED, Drd2 and Similarity

The goal of this task is to optimize molecules with $\text{QED} \geq 0.8$, $\text{Drd2} \geq 0.4$ and $\text{Similarity} \geq 0.3$.

**Molecule optimization of task 1.** In task 1, we use the data set containing 800 molecules with $\text{QED} \in [0.7, 0.8]$ as the lead molecules as in [26]. MOMO is compared against DL-based, RL-based and EC-based models (See the Methods section for the details of the baselines). The success rate (SR) is applied to evaluate the optimization performance, which is calculated as the percentage of optimized molecules with properties and similarity above their respective thresholds. **Fig. 3a** shows that QMO and MOMO achieve an overwhelming advantage in SR over other methods, up to more than 90%. MOMO still outperforming QMO by 1.91% (the results of baselines are obtained from the literature[26]). To further compare the performance of MOMO and QMO, the property distributions of the optimized molecules by QMO and MOMO in task 1 are shown in **Fig. 3b**. A dot in the figure represents an optimized molecule of the lead molecule in the test set. The horizontal axis is the value of the QED property of the optimized molecule, and the vertical axis is the Similarity of the optimized molecule to the original molecule. It indicates that MOMO optimizes more diverse and successful molecules with higher properties and similarity than QMO.

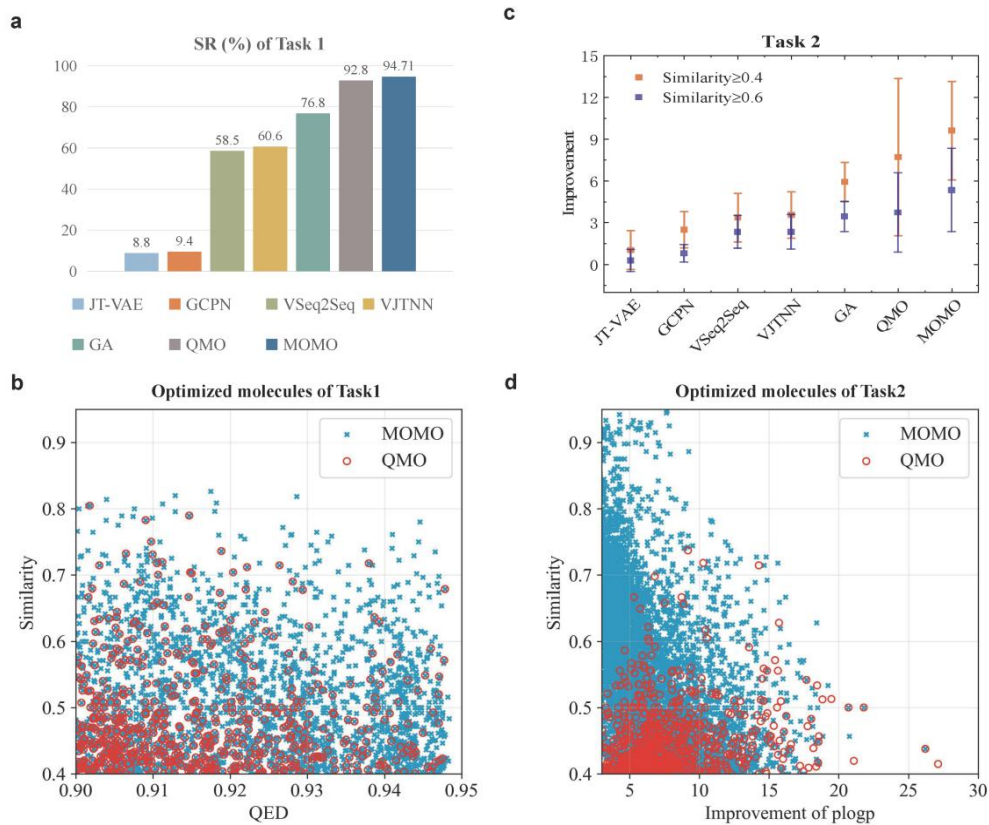

**Fig. 3. Performance evaluation of MOMO and baselines in task 1 and task 2. (a)**

Success rate of MOMO and baselines on task 1 for QED threshold of 0.9 and Similarity threshold of 0.4. **(b)** Comparison of the distribution of successful optimized molecules for the MOMO and QMO methods for task 1. **(c)** Average improvement of PlogP of MOMO and baselines on task 2 at various tanimoto similarity constraint value. **(d)** Comparison of the distribution of optimized molecules of task 2 for MOMO and QMO methods with similarity threshold of 0.4.

**Molecule optimization of task 2.** In task 2, the experiment is designed to improve the PlogP as high as possible while keeping the Similarity above threshold $\delta$. We use the dataset containing 800 low-level PlogP as in [26], and the Similarity threshold is set to 0.4 and 0.6, respectively. The results of PlogP_imp for MOMO and baselines are shown in **Fig. 3c**, where the baselines are obtained from the original literature. MOMO achieves a higher average PlogP_imp of 9.61 when $\delta = 0.4$, outperforming all baselines by at least 1.91. The standard deviation of PlogP_imp for MOMO is lower than that for QMO, suggesting that MOMO is more stable than QMO. For $\delta = 0.6$, the PlogP_imp of MOMO still outperforms the baselines, demonstrating that MOMO is robust to different Similarity thresholds, better optimizes the PlogP for molecules compared to the baseline methods. To further compare the effectiveness of MOMO and QMO, we visualize the property distribution of the optimized molecules for task 2 with a Similarity threshold 0.4 (**Fig. 3d**). It is obvious that MOMO can optimize higher PlogP_imp that maintains higher

Similarity. We notice that a optimized molecule of QMO at the bottom right corner of **Fig. 3d**, which achieves a larger PlogP_imp than MOMO. We find that this molecule is not significant because it is basically composed of a large amount of carbon without a variety of different structures.

**Molecule optimization of task 3.** To further evaluate the effectiveness of MOMO, we consider a more challenging objective combination, where three objectives are simultaneously optimized. This dataset is acquired by screening 500 molecules with QED belonging to $[0.6, 0.8]$ and $\text{PlogP} < -2$ from the datasets of tasks 1 and 2. Three baselines are compared in this experiment and the results are obtained by running the public code of the original paper. For more details of the baselines, please see the Methods section. The SR and the mean average properties (QED, PlogP_imp and Similarity) of optimized molecules are presented in **Fig. 3a**. The SR of MOMO is 82.4%, which is superior to baselines at least 32.1%. Furthermore, the mean average property values of the molecules optimized by MOMO outperform all baselines on three objectives. It shows that MOMO can optimize multiple objectives, while existing models have limited performance. To further analyze the optimized molecules, we visualize the distribution of properties of the optimized molecules with MOMO and QMO in **Fig. 3b**. To present the results more clearly, for each lead molecule, two molecules are randomly selected from the optimized molecules in MOMO and plotted in **Fig. 3b**. The three coordinates represent the values of QED, PlogP_imp, and Similarity, respectively, and the upper right corner indicates that all three properties are maximized. The results show that the optimized molecules of MOMO cover the top of the QMO, with more molecules near the upper right corner, indicating that MOMO is greater than QMO at optimizing the three objectives of molecules.

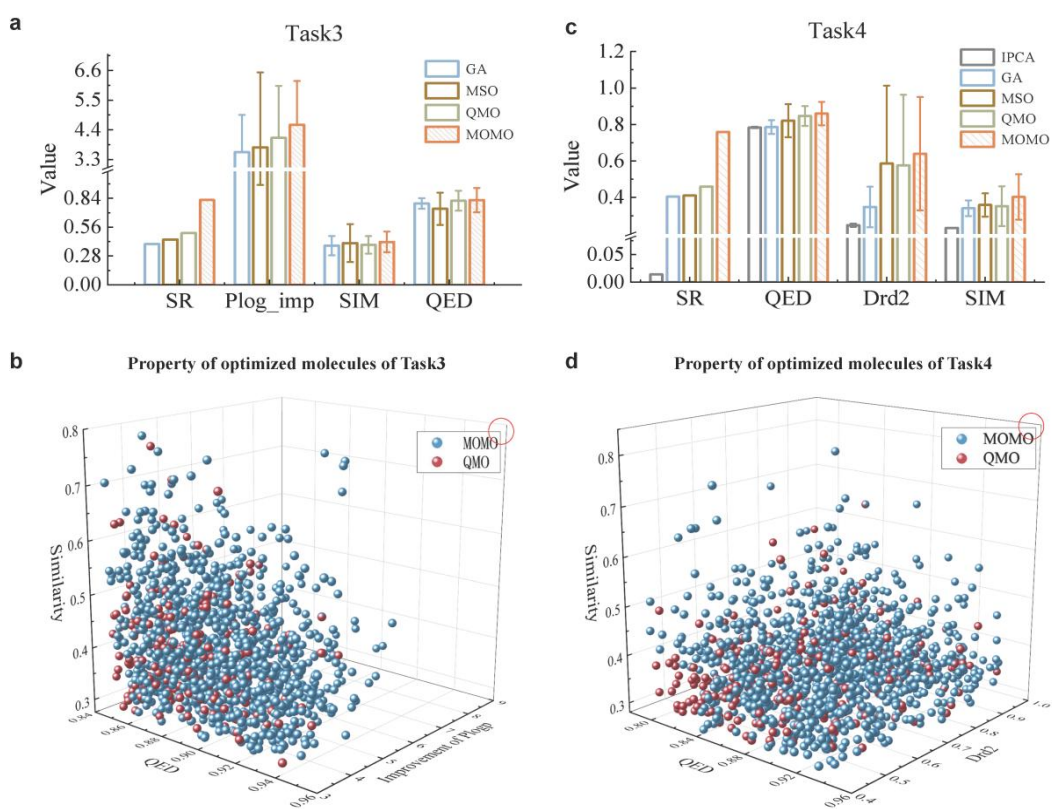

**Fig. 4. Performance evaluation of MOMO and baselines in task 3 and task 4.** Success rate and the average property values for the specified objectives are applied to evaluate the performance. **(a)** Comparison of the molecules optimized by various algorithms in task 3 for QED threshold of 0.85, PlogP_imp threshold of 3 and Similarity threshold of 0.3. **(b)** Distribution of successful optimized molecules of task 3 for MOMO and QMO methods, the top right corner is the optimal ideal point. **(c)** Comparison of the molecules optimized by various algorithms in task 4 for QED threshold of 0.8, Drd2

threshold of 0.4 and Similarity threshold of 0.3. **(d)** Distribution of optimized molecules of task 4 for MOMO and QMO methods, the top right corner is the optimal ideal point.

**Molecule optimization of task 4.** To make the results more convincing, we further consider the objective of optimizing bioactivity, which is relevant for real-world drug design. The objectives considered in task 4 are QED, Drd2 and Similarity. The dataset used in this task is a test data consisting of 780 molecules from the published article[13]. The baseline algorithms and experimental setup used for comparison in this experiment are described in the Methods section. Similar to the experiments in task 3, the performance of MOMO and baselines are evaluated using SR and mean averaged properties. The optimized SR of MOMO is 75.93%, which is at least 29.9% better than the baselines **(Fig. 4c)**, indicating that the MOMO optimizes both biological and non-biological properties of molecules. MOMO also achieves higher average property values than the baselines for all three objectives. **Fig. 4d** shows the distribution of optimized molecules for MOMO and QMO, where for MOMO, two randomly chosen optimized molecules are plotted for each lead molecule. The upper right corner of **Fig. 4d** indicates the ideal optimal molecule. It can be seen that the molecules generated by MOMO cover the top of the molecules optimized by QMO and are closer to the upper right corner, demonstrating that QMO generates molecules with low similarity when optimizing molecular properties, while MOMO can search for molecules with higher target properties that maintaining similarity.

**Visualization Analysis**

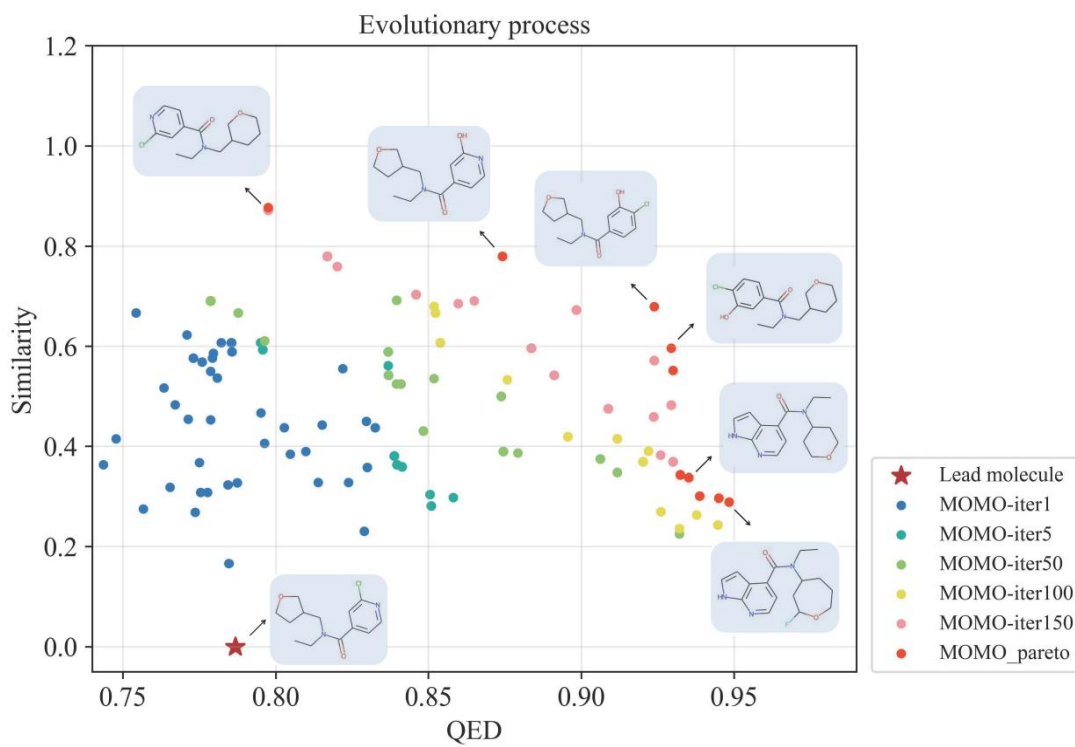

**Fig. 5. An example to demonstrate the search capability of MOMO.** The evolution of molecular properties through MOMO in task 1 is plotted. The red pentagram is the lead molecule. The dots with different colors indicate molecules in different evolutionary generations in the legend, and the final optimized molecules on the Pareto front are shown on the side of the red circles.

To analyze the search ability of MOMO, we take a molecule in task 1 as an example to show the evolutionary of molecular properties by MOMO (**Fig.5**). The pentagram in **Fig. 5** is the lead molecule, and the molecular structure is shown next to it. The dots with different colors represent the molecules in different evolutionary generations of MOMO, respectively. The red dots are the molecules on the final optimized Pareto front, and the corresponding molecular structures are plotted near the dots. As can be seen, MOMO explores multiple directions as the number of generations increases, broadly searching for molecules with larger objective values. The results indicate that MOMO can eventually search for optimized molecules with different preferences (high Similarity or high QED), and can search for successful molecules considered both Similarity and QED.

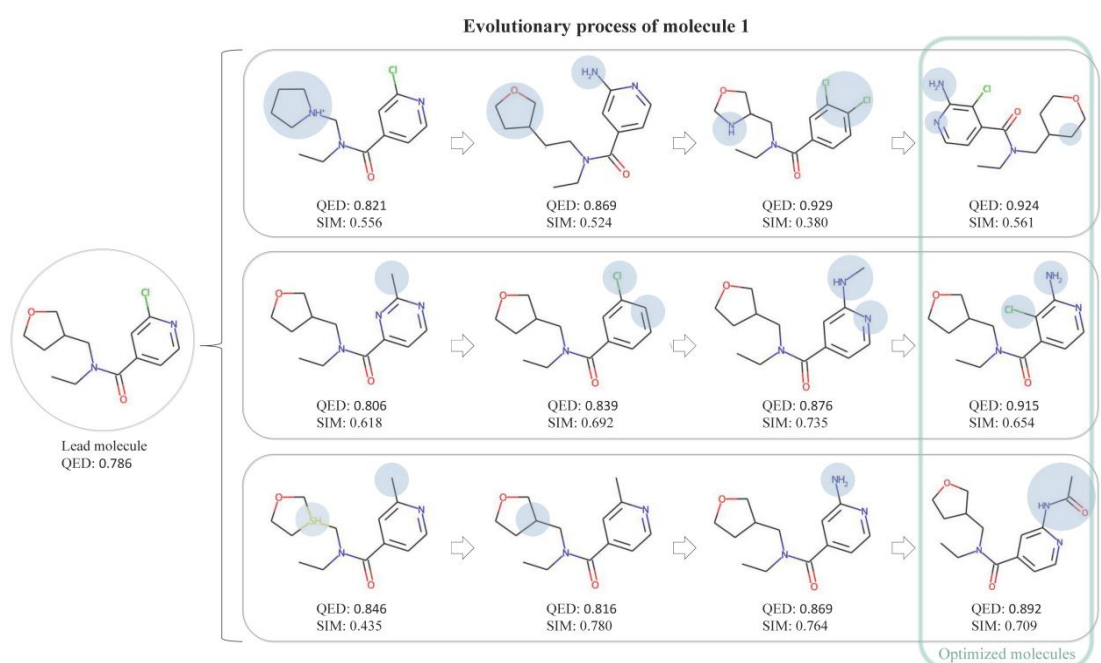

**Fig. 6. An example trajectory to demonstrate the molecular modification procedure of MOMO.** Visualization of the evolution of a molecule by MOMO in task 1. Each row shows the evolutionary of progeny molecule. From left to right, the modified substructure of the improved molecule with respect to the original one is shaded.

To further demonstrate the evolution of the molecule during the search, **Fig. 6** shows an example trajectory of a part of the evolution of a molecule in task1. Three rows represent the editorial processes of the three progenitors from the lead molecule to Pareto in different directions, respectively. The dissimilarity region of the molecule with respect to the parent molecule is shaded. The first row of molecules is searched until a successful molecule with extremely high QED and relatively low similarity is obtained. The second row generates a successful molecule with both high QED and similarity. The third row evolves to a molecule with low QED but quite similar to the lead molecule. This indicates that when a lead molecule is given, the MOMO can gradually optimize the molecule in different directions until it searches for a diverse set of molecules satisfying different objectives.

## Discussion

We formulate the molecule optimization problem as a multi-objective optimization problem and propose MOMO to purposefully evolve molecules with learned chemistry. Specifically, MOMO uses an encoder-decoder framework to construct an implicit chemical space by learning chemical knowledge from molecules, and evaluates multiple properties of molecules in the decoded molecule space to guide the evolution of molecules with desired properties in the implicit space. We demonstrated the performance of MOMO on a variety of molecule optimization tasks (**Fig. 3** and **4**), and show the search capabilities by visualizing the search process and evolutionary trajectory of MOMO (**Fig. 5** and **6**). Our multi-objective evolutionary molecule optimization framework will provide a powerful tool to improve the success rate of drug discovery by optimizing multiple objectives.

There are several improvements of MOMO compared to existing methods. First, MOMO achieves high performance across diverse multi-objective molecule optimization tasks, including the combination of drug likeness, penalized-logP, Dopamine Receptor D2 and similarity. The Pareto dominance based multi-objective evolution algorithm used in MOMO can optimize multiple objectives in parallel. Furthermore, MOMO outperforms state-of-the-art methods, including DL-based, RL-based and EC-based optimization methods, and reduces the dependence on the labeled molecules. Finally, MOMO is more stable and can explore the implicit chemical space more efficiently and smoothly to generate diverse desired molecules.

We acknowledged several limitations. Although we demonstrate the excellent performance of MOMO for optimizing multiple objectives, we find that the encoding and decoding process is time-consuming during optimization. Reducing the codec time while maintaining the accuracy of the property evaluation is challenging. Furthermore, real-world drug development commonly requires molecule optimization with more objectives in mind. Future work may additionally consider the following directions to improve our MOMO: (1) combining an agent property prediction model in latent space with a property evaluator in molecular space, to reduce the time consumed for encoding and decoding during evolution while ensuring performance; (2) consider MO of more objectives (4 objectives and more); (3) consider distributivity and diversity in molecular evolution to improve optimization performance. In summary, MOMO is an effective method to optimize the similarity and multiple properties of molecules, searching for diverse molecules with desired properties to increase the probability of drug development success.

## Methods

### Learning about chemistry

One drawback of evolutionary algorithms for MO is that they solve the problem without any prior knowledge. The lack of chemistry makes it difficult to generate valid molecules satisfying chemical constraints (such as valence bonds) when evolving in a discrete chemical space, leading to an inefficient search. To learn chemistry from a mass of molecules and guide the evolutionary process with purpose, MOMO uses a codec framework to build an implicit chemical space. The codec learns abundant intrinsic information among molecules and explores the latent space of molecules more efficiently and smoothly, further increasing the success rate of optimized molecules. Specifically, the encoder maps a discrete molecule into the latent space and obtains an embedding vector denoted by $z = E(x)$. The decoder then decodes the latent vector into discrete molecule, denoted by $x' = D(z)$. Actually,

MOMO is applicable to any codec structure with multiple molecule representation (strings[61, 62] or graphs[20, 63]) to obtain latent representations of molecules. To demonstrate the performance of the proposed framework, we combine a pre-trained cddd model[64] to obtain the implicit chemical space in this paper.

**Multi-objective evolutionary process**

MOMO uses a multi-objective evolutionary algorithm to search for molecules satisfying several objectives in an implicit chemical space. The target properties and similarity of molecules evaluated in molecule space are set as objective functions to guide the evolution process. Specifically, the lead molecule is embedded into a latent vector and an initial population is generated by perturbing the vector. Each latent vector in the population is an individual molecule. MOMO performs evolutionary operations (selection, crossover, mutation) between the vectors to generate children. Then, individuals in the parent and offspring populations are then decoded into molecular space. MOMO further calculates the values of target properties and similarity corresponding to molecules by the property evaluators (Rdkit[55] or pyTDC[56], ADMETlab[57] etc.), and directly considered the values as objectives of the multi-objective evolutionary algorithm. Moreover, non-domination rank and the reference point mechanism are used to order molecules according to their multiple target values. The ordered molecules are successively selected to the next generation of the population based on an acceptance probability, to evolution in the continuous latent space.

**Objective functions**

MO aims to generate similar molecules with desired properties by modifying the lead molecule. Several relevant properties need to be considered in MO, hence we formulate the MO problem as a multi-objective optimization problem, where each property of the molecule is set as a single objective. The goal of our MOMO framework is to find a set of Pareto solutions by solving the following problem:

$$\max_{x} [f_1(x), f_2(x), \cdots, f_m(x), f_{m+1}(x, x_0)]$$

$$s.t.\ x \in \Omega \qquad , \#(1)$$

where $f_1, f_2, \cdots, f_m$ are $m$ property predictors or evaluators, $f_{m+1}$ is a function to compute the similarity of the optimized molecule with respect to the lead molecule, $\Omega$ is the latent space.

Our multi-objective molecule optimization framework is scalable and can combine multiple objectives of the following properties or others:

(1) Comprehensive evaluation indexes for molecular drug-forming properties, such as molecular drug likeness (QED)[58], synthesizability (SA)[66], etc., which do not rely on target specific;

(2) The biological activity of molecules with disease targets;

(3) Pharmacokinetic (ADME) properties[67], like solubility, etc.

In this paper, the tanimoto molecular similarity over Morgan fingerprints[68], which has been shown to be one of the best similarity metrics and is not easily susceptible to factors such as molecular size, is applied to measure the similarity.

$$Sim(x, y) = \frac{fp(x) \cdot fp(y)}{|fp(x)|^2 + |fp(y)|^2 - fp(x) \cdot fp(y)}, \#(2)$$

where $x$, $y$ are two molecules, and $fp$ is the Morgan fingerprint of the molecule.

**Partial order of molecules**

Pareto-dominated methods are a common solution method for multi-objective evolutionary algorithms, which operate directly on the original objective values without loss of information, and are strongly applicable to any objective function[69]. To this end, in this paper, the population is updated based on Pareto domination. We employ the non-domination rank and the reference point mechanism to define the partial order of molecules according to their properties and similarity values[70]. Note that the problem is transformed into a minimization objective function in the solution process.

**Non-domination rank.** The definition of domination is as follows[69]: in a multivariate optimization problem with $m + 1$ objectives $\{f_1, f_2, \cdots, f_{m+1}\}$, the feasible solution $z_1$ dominate $z_2$ (denoted as $z_1 \prec z_2$), if $\forall i \in \{1, 2, \cdots, m + 1\}$: $f_i(D(z_1)) \leq f_i(D(z_2))$, and $\exists j \in \{1, 2, \cdots, m + 1\}$: $f_j(D(z_1)) < f_j(D(z_2))$. That is, feasible solution $z_1$ is better than $z_2$. If $z_1$ is not dominated by other solutions, it is a non-dominated solution, that is, a Pareto solution. All Pareto solutions form a Pareto-front. Thus all solutions can be sorted by Pareto ranks $F = \{F_1, F_2, \cdots\}$, rank $F_i$ dominates $F_j$ for $j > i$.

**Reference point mechanism.** We then use the reference point mechanism to define the order of individuals with the same dominant rank, which can maintain the diversity of the population by introducing widely distributed reference points. The non-dominant individuals which sparse associate with the reference point are retained[70]. The following main steps are used to select individuals: (1) Determine the reference point on the hyperplane. We adopt Das and Dennis's systematic approach[71] to generate structured reference points, which are uniformly distributed in the m-dimensional hyperplane (a total of m+1 objectives). (2) The objective of individuals in the population is adaptively normalized. (3) Associate individuals with reference points. An ideal point $f^* = (f_1^{Min}, f_2^{Min}, \cdots f_{m+1}^{Min})$ of the population is constructed and connected to each reference point to form reference lines that divide the target space symmetrically. After that, each individual is associated with the reference point on the closest reference line, and the reference point associated with solution $z_i$ is denoted as $R(z_i)$. (4) Individuals sorted. The distribution of individuals is evaluated by the niche count of the reference points, which is calculated by the number of population members associated with the reference points. The niche count of reference point $j$ is denoted as $\rho_j$. Individuals associated with reference points with small niche counts are selected first for the next generation. If more than one individual is associated with a reference point, an individual is randomly selected.

**Partial order.** The partial order of solutions is defined as follows: if feasible solution $z_1$ dominate $z_2$, i.e. $F(z_2) < F(z_2)$, $z_1$ is superior to $z_2$. If $F(z_2) = F(z_2)$, $z_1$ is superior to $z_2$ when the niche count $\rho_{R(z_1)} < \rho_{R(z_2)}$.

**Evolutionary Operations**

Evolutionary operations are defined among the latent vectors of molecules in the implicit chemical space, including selection, crossover, and mutation.

**Selection.** For the current population, we use the binary tournament selection operator[72] to select a fraction of individuals with higher fitness for evolution. Specifically, two individuals are randomly sampled from the population, and the one with larger fitness is selected to the evolve population. This process is repeated several times until the size of the population reaches that of the original population.

**Crossover.** For each individual $z_1$, another individual $z_2$ is randomly selected to perform crossover and generate a new offspring. The blending linear crossover operator is applied to produce new children[73]:

$$\begin{cases} z'_1 = z_1 + c_1(z_2 - z_1), \ c_1 = -d + (1 + 2d)\, u_1 \\ z'_2 = z_1 + c_2(z_2 - z_1), \ c_2 = -d + (1 + 2d)\, u_2 \end{cases}, \#(3)$$

where $u_1, u_2 \sim \text{Uniform}[0, 1]$, and $d \geq 0$ is a parameter that controls whether the exploration space is interpolated or extrapolated.

**Mutation.** Individuals generated by crossover continue to mutate with a certain probability ($p_m = 0.5$). A random value $r$ is sampled for individual $z$ ($D$-dimensional vector), if $r < p_m$, an integer $c$ is randomly drawn from $\{1, \cdots, D\}$, and the $c$-th position is replaced with a random value drawn from the standard Gaussian distribution.

**Update population**

Excellent individuals are maintained in a new population to guarantee that the population evolves towards the target. In this way, previous population ($Z$) and children ($Z'$) are merged ($\tau = Z \cup Z'$) and decoded into SMILES to evaluate their properties and similarity. Desired molecules are retained for the next population based on their partial order with the reception probability. The specific process of forming the new population is as follows.

All individuals in $\tau$ are first sorted by the non-domination rank and then by the reference point mechanism. In addition, we design the acceptance probability to update population, which can maintain the diversity of molecules in the population and prevent fall into some local optimum regions. So that the top of the ranked molecules (high objective values) have a certain probability to enter the next generation or may not be accepted. The acceptance probability is calculated as follow:

$$p_a = e^{-\frac{1}{t} \times \beta}, \#(4)$$

where $t$ represents the current index of evolution epoch, $p_a$ is increasing with the iteration of $t$, and $\beta$ controls the incremental speed, which is set as 0.3 in experiments. The acceptance probability is applied such that molecules with low objective values at early evolutionary times are still likely to enter the next generation to emphasize diversity in the population, while molecules with higher objective values are accepted at later evolutionary times to emphasize the properties and similarity.

Finally, each molecule is chosen sequentially to construct a new population. Starting from the top molecule, a random value $r$ is sampled, if $r < p_a$, the molecule is accepted. The next molecule is then processed until the size of the new population is equal to $P$.

**Experimental settings**

In our experiments, we input the lead molecules in the dataset, run MOMO based on the set parameters, and report the results for the Pareto-front molecules found in the last generation of the evolved population. Note that invalid molecules are discarded in the search process. In the following, we specify the datasets used in the experiments, the comparison models, the evaluation metrics, and the experimental parameter sets.

**Datesets.** We use four different datasets for four tasks. The QED dataset for task 1 is provided by Jin et al.[20], containing 800 molecules with $\text{QED} \in [0.7, 0.8]$ selected from ZINC test[74]. The PlogP dataset used for task 2 contains 800 low-level PlogP as in Jin et al[20]. To acquire the dataset used in task 3, we merge the QED dateset and PlogP dateset and screening 500 unique molecules with QED belonging to [0.6, 0.8] and $\text{PlogP} <- 2$. For task 4, the used dataset is a test data consisting of 780 molecules with $\text{Drd2} < 0.05$ and $\text{QED} \in [0.7, 0.8]$ from the published article[13,21].

**Baselines.** In task 1 and task 2, we compare MOMO with three different types of popular baselines, which are the DL-based methods (JTVAE[20], VSeqtoSeq[14], VJTNN[21], QMO[26]), the RL-based method (GCPN[44]) and the EC-based method (GA[43]).

In task 3 and task 4, due to the increased difficulty of the optimization tasks, some of the above models perform poorly on tasks where multiple properties are optimized simultaneously, or some of them are difficult to retrain for different tasks. Three methods are re-selected as baselines in tasks 3 and 4, containing GA[43], MSO[15] and QMO[26]. Note that all the multiple objectives in the three models are combined into a single objective function. In addition, in task 4 we also compared a deep learning model based on data domain transformation rather than combine objectives (IPCA[13]), which requires the definition of data domains with high and low properties. However, the PlogP metric is undecidable to define domains, and therefore we did not perform the comparison in task 3 to prevent biased results.

**Benchmark.** Several metrics are used to evaluate our tasks: (1) Success rate (SR) is defined as the percentage of optimized molecules which similarity and property values above their respective thresholds. (2) The average objective value is calculated by the average property (such as QED, PlogP, DRD2) or similarity value of the optimized molecules. (3) The average improvement of property is the average property value of the optimized molecules subtracted from the average value of the lead molecules.

**Parameter setting.** For the QED and Similarity optimization task 1, we use $P = 100$, $T = 100$, $d = 0.25$, $p_c = 1$, $p_m = 0.5$ in MOMO, and adopt multiple restarts of the initial population to improve the performance. The reported results are the best among the 10 restarts. For the baseline GA, the population size is 100 and the number of iterations is 50. The QED value is set as the objective function and Similarity is used as the penalty with a threshold of 0.4. The results for the other comparison models are obtained from the original paper.

For the PlogP and Similarity optimization task 2: the model parameters are set to $P = 100$, $T = 200$, $d = 0.25$, $p_c = 1$, $p_m = 0.5$ in MOMO. The experimental results for the comparison model are those of the original paper.

For both task 3 and task 4, no restarts are considered in all methods in order to fairly compare all experiments and take into account the computational costs. We use $P = 100$, $T = 200$, $d = 0.25$, $p_c = 1$, $p_m = 0.5$ in MOMO. Since multiple objectives are combined into a single objective function in comparison models, to make the results more reliable, several sets of weights are generated for the experiments by applying the method of generating uniform weights, and the corresponding results for the parameters with the best final results are shown. For the baselines, the population size is 500 with 100 iterations in GA, the number of particles is 200 with 200 iterations in MSO, and for QMO, the sample size is 200 with 100 runs. The IPCA results used for comparison in task 4 are taken from the original paper.

## Data Availability

The datasets used in this project will be update and available at auther's Github.

## Code Availability

All of the codes will be update and available at auther's Github.

## Acknowledgments

**Funding**:

This work was supported by the National Key Research and Development Program of China (2021YFE0102100), the Natural Science Foundation of China under grants (62172002, 62122025, 61872309, U19A2064).

## Author Contributions Statement

All authors devised the research project. X. X. developed the codes, X. X. and Y. S. implemented experiments and analyzed the results. X. X., Y. S. and X. Z. wrote and critically revised the manuscript.

## Competing Interests Statement

The authors have declared no competing interests.

# References

1. De Rycker, M., Baragaña, B., Duce, S.L. & Gilbert, I.H. Challenges and recent progress in drug discovery for tropical diseases. *Nature* **559**, 498-506 (2018).
2. Lowe, D. The latest on drug failure and approval rates. *Sci. Transl. Med.* 2 (2019).
3. Schneider, G. Automating drug discovery. *Nat. Rev. Drug Discovery* **17**, 97-113 (2018).
4. Sheng, C.Q., Li, J. Structural optimization of drugs: Design strategies and empirical rules. *Chemical Industry Press*. Beijing, 7 (2017).
5. Chen, Z., Min, M. R., Parthasarathy, S., & Ning, X. A deep generative model for molecule optimization via one fragment modification. *Nat. Mach. Intell.* **3**, 1040-1049 (2021).
6. Hsu, H. H., Hsu, Y. C., Chang, L. J., & Yang, J. M. An integrated approach with new strategies for QSAR models and lead optimization. *BMC genomics.* **18**, 1-9 (2017).
7. Zhavoronkov, A. Artificial intelligence for drug discovery, biomarker development, and generation of novel chemistry. *Mol. Pharm.* **15**, 4311-4313 (2018).
8. Graff, D. E., Shakhnovich, E. I., & Coley, C. W. Accelerating high-throughput virtual screening through molecular pool-based active learning. *Chem. Sci.* **12**, 7866-7881 (2021).
9. Polishchuk, P. G., Madzhidov, T. I. & Varnek, A. Estimation of the size of drug-like chemical space based on GDB-17 data. *J. Comput. Aided Mol. Des.* **27**, 675–679 (2013).
10. Gao, W., Fu, T., Sun, J., Coley, C. W. Sample Efficiency Matters: A Benchmark for Practical Molecular Optimization. 2022; http://arxiv.org/abs/2206.12411.
11. Vamathevan, J. et al. Applications of machine learning in drug discovery and development. *Nat. Rev. Drug. Discov.* **18**, 463-477 (2019).
12. Jiménez-Luna, J., Grisoni, F., & Schneider, G. Drug discovery with explainable artificial intelligence. *Nat. Mach. Intell.* **2**, 573-584 (2020).
13. Barshatski, G., Nordon, G., & Radinsky, K. Multi-Property Molecular Optimization using an Integrated Poly-Cycle Architecture. In *Proc. 30th ACM International Conference on Information & Knowledge Management* 3727-3736 (2021).
14. Bahdanau, D., Cho, K. & Bengio, Y. Neural machine translation by jointly learning to align and translate. *In Proc. International Conference on Learning Representations* (2015).
15. Winter, R., Montanari, F., Steffen, A., Briem, H., Noé, F., & Clevert, D. A. Efficient multi-objective molecular optimization in a continuous latent

space. *Chem. Sci.* **10**, 8016-8024 (2019).
16. Xie, Y., Shi, C., Zhou, H., Yang, Y., Zhang, W., Yu, Y., & Li, L. Mars: Markov molecular sampling for multi-objective drug discovery. In *Proc. International Conference on Learning Representations* (2021).
17. Grantham, K., Mukaidaisi, M., Ooi, H. K., Ghaemi, M. S., Tchagang, A., & Li, Y. Deep Evolutionary Learning for Molecular Design. *IEEE Comput. Intell. M.* **17**, 14-28 (2022).
18. Sousa, T., Correia, J., Pereira, V., & Rocha, M. Combining Multi-objective Evolutionary Algorithms with Deep Generative Models Towards Focused Molecular Design. In *International Conference on the Applications of Evolutionary Computation* (Part of EvoStar) 81-96 (2021).
19. Gómez-Bombarelli, R. et al. Automatic chemical design using a data-driven continuous representation of molecules. *ACS Cent. Sci.* **4**, 268–276 (2018).
20. Jin, W., Barzilay, R., & Jaakkola, T. Junction tree variational autoencoder for molecular graph generation. In *Proc. International Conference on Machine Learning* 2323–2332 (PMLR, 2018).
21. Jin, W., Yang, K., Barzilay, R., & Jaakkola, T. Learning multimodal graph-to-graph translation for molecule optimization. In *Proc. International Conference on Learning Representations* (2019).
22. Yu, J., Xu, T., Rong, Y., Huang, J., & He, R. Structure-aware conditional variational auto-encoder for constrained molecule optimization. *Pattern Recognition*. **126**, 108581 (2022).
23. Lee, M., Min, K. MGCVAE: Multi-Objective Inverse Design via Molecular Graph Conditional Variational Autoencoder. *J. Chem. Inf. Model*. **62**, 2943–2950 (2022).
24. He, J. et al. Molecular optimization by capturing chemist's intuition using deep neural networks. *J. Cheminform.* **13**, 1-17 (2021).
25. Barshatski, G., & Radinsky, K. Unpaired Generative Molecule-to-Molecule Translation for Lead Optimization. In *Proc. 27th ACM SIGKDD Conference on Knowledge Discovery & Data Mining* 2554-2564 (2021).
26. Hoffman, S. C., Chenthamarakshan, V., Wadhawan, K., Chen, P. Y., & Das, P. Optimizing molecules using efficient queries from property evaluations. *Nat. Mach. Intell.* **4**, 21-31 (2022).
27. He, J. et al. Transformer-Based Molecular Optimization Beyond Matched Molecular Pairs. *J. Cheminform.* **14**, 18 (2022).
28. Sanchez-Lengeling, B., Outeiral, C., Guimaraes, G. L. & Aspuru-Guzik, A. Optimizing distributions over molecular space. An objective-reinforced generative adversarial network for inverse-design chemistry (organic). Preprint at https://doi.org/10.26434/chemrxiv.5309668.v3 (2017).

29. De Cao, N., & Kipf, T. MolGAN: An implicit generative model for small molecular graphs. *ICML 2018 workshop on Theoretical Foundations and Applications of Deep Generative Models* (2018).
30. Maziarka, Ł., Pocha, A., Kaczmarczyk, J., Rataj, K., Danel, T., & Warchoł, M. Mol-CycleGAN: a generative model for molecular optimization. *J. Cheminform.* **12**, 1-18 (2020).
31. Fu, T., Xiao, C., Li, X., Glass, L. M., & Sun, J. Mimosa: Multi-constraint molecule sampling for molecule optimization. In *Proc. AAAI Conference on Artificial Intelligence* 125-133 (AAAI, 2021).
32. Ji, C., Zheng, Y., Wang, R., Cai, Y., & Wu, H. Graph Polish: a novel graph generation paradigm for molecular optimization. *IEEE Trans. Neural. Netw. Learn. Syst* (2021).
33. Fu, T., Xiao, C. & Sun, J. CORE: automatic molecule optimization using copy & refine strategy. In *Proc. AAAI Conference on Artificial Intelligence* 638–645 (AAAI, 2020).
34. Olivecrona, M., Blaschke, T., Engkvist, O., & Chen, H. Molecular de-novo design through deep reinforcement learning. *J. Cheminform.* **9**, 1-14 (2017).
35. Popova, M., Isayev, O., & Tropsha, A. Deep reinforcement learning for de novo drug design. *Sci. Adv.* **4**, 7885 (2018).
36. Zhou, Z., Kearnes, S., Li, L., Zare, R. N. & Riley, P. Optimization of molecules via deep reinforcement learning. *Sci. Rep.* **9**, 10752 (2019).
37. Jin, W., Barzilay, R., & Jaakkola, T. Multi-objective molecule generation using interpretable substructures. In *International conference on machine learning* 4849-4859 (PMLR, 2020).
38. Jensen, J. H. A graph-based genetic algorithm and generative model/ Monte Carlo tree search for the exploration of chemical space. *Chem. Sci.* **10**, 3567-3572 (2019).
39. Kwon, Y., & Lee, J. MolFinder: an evolutionary algorithm for the global optimization of molecular properties and the extensive exploration of chemical space using SMILES. *J. Cheminform.* **13**, 1-14 (2021).
40. Nigam, A., Pollice, R., Krenn, M., dos Passos Gomes, G., & Aspuru-Guzik, A. Beyond generative models: superfast traversal, optimization, novelty, exploration and discovery (STONED) algorithm for molecules using SELFIES. *Chem. Sci.* **12**, 7079-7090 (2021).
41. Kwon, Y., & Lee, J. MolFinder: an evolutionary algorithm for the global optimization of molecular properties and the extensive exploration of chemical space using SMILES. *J. Cheminform.* **13**, 1-14 (2021).
42. Zhou, A., Qu, B. Y., Li, H., Zhao, S. Z., Suganthan, P. N., & Zhang, Q. Multiobjective evolutionary algorithms: A survey of the state of the art. *Swarm. Evol. Comput.* **1**, 32-49 (2011).
43. Nigam, A., Friederich, P., Krenn, M., & Aspuru-Guzik, A. Augmenting genetic algorithms with deep neural networks for exploring the chemical space. *International Conference on Learning Representations* (2020).
44. You, J., Liu, B., Ying, Z., Pande, V. & Leskovec, J. Graph convolutional policy network for goal-directed molecular graph generation. In *Advances in Neural Information Processing Systems* 6410–6421 (NIPS, 2018).
45. Gaikwad, R., & Lakshmanan, R. Study of Evolutionary Algorithms for Multi-objective Optimization. *SN Computer Science*. **3**, 1-7 (2022)..
46. Tian, Y., Si, L., Zhang, X., Cheng, R., He, C., Tan, K. C., & Jin, Y. Evolutionary large-scale multi-objective optimization: A survey. *ACM Computing Surveys (CSUR)*. **54**, 1-34 (2021).
47. Deng, W., Zhang, X., Zhou, Y., Liu, Y., Zhou, X., Chen, H., & Zhao, H. An enhanced fast non-dominated solution sorting genetic algorithm for multi-objective problems. *Inform. Sciences.* **585**, 441-453 (2022).
48. Tian, Y., Cheng, R., Zhang, X., & Jin, Y. PlatEMO: A MATLAB platform for evolutionary multi-objective optimization [educational forum]. *IEEE Comput. Intell. M.* **12**, 73-87 (2017).
49. Tian, Y., Su, X., Su, Y., & Zhang, X. EMODMI: A multi-objective optimization based method to identify disease modules. *IEEE Transactions on Emerging Topics in Computational Intelligence.* **5**, 570-582 (2020).
50. S. V., S. S.; Law, J. N.; Tripp, C. E.; Duplyakin, D.; Skordilis, E.; Biagioni, D.; Paton, R. S.; St. John, P. C. Multi-Objective Goal-Directed Optimization of De Novo Stable Organic Radicals for Aqueous Redox Flow Batteries. *Nat. Mach. Intell.* **4**, 720–730 (2022).
51. Schneider, P. et al. Rethinking drug design in the artificial intelligence era. *Nat. Rev. Drug. Discov.* **19**, 353-364 (2020).
52. Bilodeau, C., Jin, W., Jaakkola, T., Barzilay, R., & Jensen, K. F. Generative models for molecular discovery: Recent advances and challenges. *Wiley Interdisciplinary Reviews: Computational Molecular Science* 1608 (2022).
53. Fromer, J. C., & Coley, C. W. Computer-Aided Multi-Objective Optimization in Small Molecule Discovery. *arXiv preprint arXiv:2210. 07209* (2022).
54. Verhellen, J. Graph-Based Molecular Pareto Optimisation. *Chem. Sci.* (2022).
55. Bento, A. P. et al. An open source chemical structure curation pipeline using RDKit. *J. Cheminform.* **12**, 1-16 (2020).
56. Huang, K. et al. Therapeutics data commons: Machine learning datasets and tasks for drug discovery and development. *NeurIPS 2021 Datasets and Benchmarks* (2021).
57. Xiong, G. et al. ADMETlab 2.0: an integrated online platform for accurate and comprehensive predictions of ADMET properties. *Nucleic Acids Research*. **49**, W5-W14 (2021).

58. Bickerton, G. R., Paolini, G. V., Besnard, J., Muresan, S. & Hopkins, A. L. Quantifying the chemical beauty of drugs. *Nat. Chem.* **4**, 90–98 (2012).
59. Duan, J., Wainwright, M. S., Comeron, J. M., Saitou, N., Sanders, A. R., Gelernter, J., & Gejman, P. V. Synonymous mutations in the human dopamine receptor D2 (DRD2) affect mRNA stability and synthesis of the receptor. *Hum. Mol. Genet.* **12**, 205-216 (2003).
60. Bajusz, D., Rácz, A., & Héberger, K. Why is Tanimoto index an appropriate choice for fingerprint-based similarity calculations?. *J. cheminform.* **7**, 1-13 (2015).
61. Weininger, D. SMILES, a chemical language and information system. 1. Introduction to methodology and encoding rules. *J. Chem. Inf. Comput. Sci.* **28**, 31–36 (1988).
62. Krenn, M., Häse, F., Nigam, A., Friederich, P., & Aspuru-Guzik, A. Self-referencing embedded strings (SELFIES): A 100% robust molecular string representation. *Machine Learning: Science and Technology*. **1**, 045024 (2020).
63. Zeng, X., Xiang, H., Yu, L., Wang, J., Li, K., Nussinov, R., & Cheng, F. Accurate prediction of molecular targets using a self-supervised image representation learning framework. *Res. Sq.* rs-3(2022).
64. Winter, R., Montanari, F., Noé, F., & Clevert, D. A. Learning continuous and data-driven molecular descriptors by translating equivalent chemical representations. *Chem. Sci.* **10**, 1692-1701 (2019).
65. Deb, K., & Jain, H. An evolutionary many-objective optimization algorithm using reference-point based nondominated sorting approach, part I: solving problems with box constrains. *IEEE T. Evolut. Comput.* **18**, 577-601 (2013).
66. Gao, W., & Coley, C. W. The synthesizability of molecules proposed by generative models. *J. Chem. Inf. Model.* **60**, 5714-5723 (2020).
67. Di, L., & Kerns, E. Drug-like properties: concepts, structure design and methods from ADME to toxicity optimization. *Academic press* (2015).
68. Rogers, D., & Hahn, M. Extended-connectivityfingerprints. *J. Chem. Inf. Model.* **50**, 742-754 (2010).
69. Deb, K., Pratap, A., Agarwal, S., & Meyarivan, T. A. M. T. A fast and elitist multiobjective genetic algorithm: NSGA-II. *IEEE T. Evolut. Comput.* **6**, 182-197 (2002).
70. Deb, K., & Jain, H. An evolutionary many-objective optimization algorithm using reference-point-based nondominated sorting approach, part I: solving problems with box constraints. *IEEE T. Evolut. Comput.* **18**, 577-601 (2013).
71. Das, I., & Dennis, J. E. Normal-boundary intersection: A new method for generating the Pareto surface in nonlinear multicriteria optimization problems. *SIAM journal on optimization.* **8,** 631-657 (1998).

72. Razali, N. M., & Geraghty, J. Genetic algorithm performance with different selection strategies in solving TSP. In *Proc. world congress on engineering.* Hong Kong, China: International Association of Engineers 1-6 (2011).
73. Takahashi, M., & Kita, H. A crossover operator using independent component analysis for real-coded genetic algorithms. In *Proc. 2001 Congress on Evolutionary Computation* 643-649 (IEEE, 2001).
74. Sterling, T., & Irwin, J. J. ZINC 15–ligand discovery for everyone. *J. Chem. Inf. Model.* **55**, 2324-2337 (2015).